\documentclass{elsart}
\newtheorem{myth}{Theorem}

\newtheorem{myle}{Lemma}

\newtheorem{mycor}{Corollary}

\usepackage{amsfonts, graphicx}
\begin{document}
\runauthor{Karagiorgos and Poulakis}
\begin{frontmatter}

\title{Efficient algorithms for the basis of finite Abelian groups}

\author[Athens]{Gregory Karagiorgos}
\author[Thessaloniki]{and Dimitrios Poulakis \thanksref{Someone}}

\address[Athens]{Department of Informatics and Telecommunications \\
 University of Athens \\Panepistimioupolis 157 84, Athens, Greece}
\address[Thessaloniki]{Department of Mathematics,\\
 Aristotle University of Thessaloniki, \\
 Thessaloniki 54124, Greece}

\thanks[Someone]{Email: greg@di.uoa.gr, poulakis@math.auth.gr}

\begin{abstract}
Let $G$ be a finite abelian group $G$ with $N$ elements. In this
paper we give  a $O(N)$ time algorithm for computing a basis of
$G$. Furthermore, we obtain an algorithm for computing a basis
from a generating system of $G$ with $M$ elements having time
complexity $O(M\sum_{p|N} e(p)\lceil p^{1/2}\rceil^{\mu(p)})$,
where $p$ runs over all the prime divisors of $N$, and $p^{e(p)}$,
$\mu(p)$ are the exponent and the number of cyclic groups which
are direct factors of the $p$-primary component of $G$,
respectively. In case where $G$ is a cyclic group having a generating system with $M$
elements, a $O(MN^{\epsilon})$ time algorithm for the computation of a basis of $G$ is obtained.
\end{abstract}

\end{frontmatter}

\section{Introduction}

In recent years, interest in studying finite abelian groups has
raised due to the increasing significant of its relationship with
public key cryptography, quantum computing and error-correcting
codes. Abelian groups as the groups $\Zset_n^{*}$ of invertible
elements of $\Zset_n$, the multiplicative groups of finite fields,
the groups of elements of elliptic curves over finite fields, the
class groups of quadratic fields and others have been used for the
designation of public key cryptosystems \cite{Koblitz}. On the
other hand, in quantum computing, the famous hidden subgroup
problem in case of a finite abelian group has been solved by a
polynomial time quantum algorithm \cite{Cheung, Kitaev, Lomont}.
The Shor's algorithm to factorize integers is one very important
special case \cite{Shor}. Recently, an interesting application of
finite abelian groups has been given in the construction  of
efficient error correcting codes \cite{Dinur}.

The problem of determining if two groups are isomorphic is one of
the fundamental problems in group theory and computation. The
group isomorphism problem is related to the graph isomorphism
problem. Interesting results have been obtained for the case of
abelian groups \cite{Vikas, Kavitha2, Garzon}. Recently, Kavitha
\cite{Kavitha} proved that group isomorphism for abelian groups
with $N$ elements can be determined in $O(N)$ time.

Let $(G,+)$ be a finite abelian group. Let $H_1, \ldots , H_r$ be
subgroups of $G$. The set of elements $x_1+\cdots + x_r$, where
$x_i \in H_i$   $(i =1,\ldots,r)$ is denoted by $H_1+\cdots + H_r$
and is a subgroup of $G$  called the {\em sum} of $H_1, \ldots ,
H_r$. It is called {\em direct sum} if for every $i =1,\ldots,r$
we have $H_i \cap (H_1+\cdots +H_{i-1}+H_{i+1}+\cdots +H_r) =
\{0\}.$  In this case we write $H_1\oplus\cdots \oplus H_r$. If $x
\in G$, then the set $<x>$ of elements $ax$ where $a  \in \Zset$
is a subgroup of $G$ called the  {\em cyclic group} generated by
$x$. Let $S \subseteq G$. The group $<S> = \sum_{x\in S} <x>$ is
called  the  {\em group generated} by $S$.  In case where $G =
<S>$, the set $S$ is called a {\em generating system} for $G$.
Suppose now that $G$ has $N$ elements and  $N = p_1^{a_1}\cdots
p_k^{a_k}$ is the prime factorization of $N$. It is well known
that $G = G(p_1)\oplus \cdots \oplus G(p_k)$, where $G(p_i)$ is a
subgroup of $G$ of order $p_i^{a_i}$ $(i=1,\ldots, k)$
\cite[Theorem 16, page 96]{Ledermann} and is called the $p_i$-{\em
primary component} of $G$. Furthermore, for every $i =1,\ldots,k$,
$G(p_i)$ can be decomposed to a direct sum of cyclic groups
$<x_{i,j}>$ $(j=1,\ldots, \mu(p_i))$ with prime-power order. The set
of elements $x_{i,j}$ $(i= 1,\ldots, k, \ j = 1,\ldots , \mu(p_i))$
is called a {\em basis} of $G$. The smallest prime power
$p_i^{e(p_i)}$ such that $p_i^{e(p_i)}x = 0$, for every $x\in G(p_i)$
is called the {\it exponent} of $G(p_i)$.

 The elements of a basis  with its orders fully determine the structure
 of a finite abelian group. Thus, an algorithm for finding a basis
of a finite abelian group can be easily converted to an algorithm
of checking the isomorphism of two such groups. Therefore, the
development of efficient algorithms for the determination of the
basis of a finite abelian group has fundamental significance in
all the above applications.

In \cite{Chen}, Chen  gave an $O(N^2)$ time algorithm  for finding
a basis of a finite abelian group $G$. Recently, in \cite{Chen2},
Chen and Fu showed an $O(N)$ time  algorithm  for this task. In
case where $G$ is represented  by an explicit set of $M$
generators, a $O(MN^{1/2+\epsilon})$ time algorithm is given by
Iliopoulos \cite{Iliopoulos} and $O(MN^{1/2})$ time algorithms are
obtained by Teske \cite{Teske2} and by Buchmann and Schmidt
\cite{Buchmann}. When $G$ is represented by a set of defining
relations that is associated with an integer matrix $M(G)$, the
computation of the structure of $G$ can be reduced to computing
the Smith Normal Form of $M(G)$. A such approach can be found in
\cite{Iliopoulos3}. Finally, in \cite{Borges}, an algorithm is
given for computing the structure of $G$ based on Gr\"{o}bner
bases techniques.

In this paper we give a simple deterministic algorithm for the
computation of a basis of a finite abelian group. More precisely,
we prove the following theorem.

\begin{myth}
Let $G$ be an  abelian group $G$ with $N$ elements. There is an
$O(N)$ time algorithm for computing a basis of $G$.
\end{myth}

An immediate consequence of the above theorem is the following
corollary.

\begin{mycor}
Group isomorphism for abelian groups with $N$ elements can be
determined in $O(N)$ time.
\end{mycor}

Adapting our method in case where a generating system of $G$ is
known,  we obtain the following theorem.

\begin{myth}
Let $G$ be an  abelian group with $N$ elements having a generating
system with $M$ elements. Let $N = p_1^{a_1}\cdots p_k^{a_k}$ be
the prime factorization of $N$ and $\mu(p_i)$, $\e(p_i)$ as above.
 Then, there is an algorithm for computing a basis of $G$ with
 time complexity $O(M\sum_{i=1}^{k} e(p_i)\lceil p_i^{1/2}\rceil^{\mu(p_i)})$ and
$\Omega(\sum_{i=1}^{k}p_i^{(\mu(p_i)-1)/2})$.
\end{myth}

In the special case where the group $G$ is cyclic, we have  the
following better result.

\begin{myth}
Let $G$ be a  cyclic group with $N$ elements having a generating
system with $M$ elements. Then, there is an algorithm for
computing a basis of $G$ with
 time complexity $O(MN^{\epsilon})$, where $\epsilon$ is arbitrary small.
\end{myth}

The algorithm  of Chen and Fu \cite{Chen2}, which computes a basis of $G$,
 is relied on an algorithm of Kavitha \cite{Kavitha}
for computing the orders of all elements in $G$. Our approach is
completely different. The basic idea of our paper is the
construction of bases of successively larger subgroups of the
$p$-components $G(p)$ of $G$, using an algorithm of Beynon and
Iliopoulos \cite{Beynon} and an algorithm of Teske \cite{Teske},
until a basis of $G(p)$ is obtained. Note that \cite{Iliopoulos2}
yields that the time complexity in Theorem 1 and 2 is reasonable.

We assume that $G$ is given by its multiplication table.
 This is equivalent to the multiplication oracle where each
 group operation can be performed in constant time.
Since the expected running time that an integer $N$ can be
factorized into product of prime numbers is subexponential (see
\cite[Chapter 19]{Gathen} and \cite[Section 2.3]{Yan}), we assume
that the prime factorization of $N$, which is the size of abelian
group, is known. To measure the running time of  our algorithm we
count  the number of numerical operations, comparisons and the
number of group operations.

The paper is organized as follows. In section 2 we give some
auxiliary results on finite abelian groups which are necessary for
the presentation of our algorithms. In section 3, 4  and 5 we give the algorithms which
correspond to
 Theorems 1, 2 and 3.

\section{Auxiliary Results}

In this section  we give some results  very useful for the design
of our algorithms. We denote by $|A|$ the cardinality of a finite
set $A$. If $x$ is a real number, then we denote by $\lceil x
\rceil$, as usually, the smallest integer $z$ such that $x \leq
z$. Let $(G,+)$ be a finite abelian group. For $x \in G$, the {\em
order} of $G$, denoted by ord$(x)$, is the smallest positive
integer $\mu$ such that ${\mu}x = 0$ (where $0$ is the identity
element of $G$).  If $S=\{x_1,\ldots,x_k\} \subseteq G$, then we
also write $<S> = <x_1,\ldots,x_n>$.

\begin{myle}
 If $x\in G$ has order $m =
p_1^{b_1}\cdots p_k^{b_k}$, then for $i = 1,\ldots, k$ the element
$x_i = (m/p_i^{b_i})x$ has order $p_i^{b_i}$ and we have
$$<x> =<x_1> \oplus \cdots \oplus <x_k>.$$
\end{myle}
\begin{pf}
See \cite[page 96]{Ledermann}. \qed
\end{pf}

Let  $|G| = N$.  We assume that the prime factorization of $N =
p_1^{a_1}\cdots p_k^{a_k}$ is given. The positive divisors of $N$
are the integers $d$ given by $d = p_1^{b_1}\cdots p_k^{b_k}$,
where $0 \leq b_i \leq a_i$ $(i = 1,\ldots,k)$.  We denote by
$\tau(N)$ the number of positive divisors of $N$. Put
$$B =
\{(b_1,\ldots,b_k)\in {\Zset}^k/ \  0 \leq b_i \leq a_i, \ \forall
i = 1,\ldots,k \}.$$ The following algorithm provide us with the
positive divisors of $N$.

\ \\
{\bf DIVISORS}
\ \\
{  \sl Input:} $N$ and its prime factorization $N= p_1^{a_1}\cdots
p_k^{a_k}$.
\ \\
{ \sl Output:} $D(N)=\{(d_i,(b_{i,1},\ldots,b_{i,k}))/i = 1,
\ldots, d_{\tau(N)}\}$, where $d_1< \cdots < d_{\tau(N)}$ are the
positive divisors of $N$  sorting in increasing order  and
$(b_{i,1},\ldots,b_{i,k})\in B$ with $d_i = p_1^{b_{i,1}}\cdots
p_k^{b_{i,k}}$.

\begin{enumerate}
\item For $(b_1,\ldots,b_k) \in B$
\begin{enumerate}
\item Compute  $d(b_1,\ldots,b_k) = p_1^{b_1}\cdots p_k^{b_k}$.

\item Store $d(b_1,\ldots,b_k)$ and the corresponding $k$-tuple
$(b_1, \ldots, b_k)$ in binary search tree.

\end{enumerate}
\item Using the binary tree sort,
sort the divisors of $N$  in increasing order $d_1< \cdots <
d_{\tau(N)}$.
\end{enumerate}

The computation of each divisor $d$ requires $O(({\log}d)^2)$
arithmetic operations. Thus the time of computation of all
positive divisors of $N$ is $O(\tau(N) ({\log}N)^2)$. The time
complexity of binary search tree and binary tree sort algorithms
is $O(\tau(N))$. Further, by \cite[Theorem 315, page 260]{Hardy},
we have $\tau(N) = O(N^{\epsilon})$, for every positive
$\epsilon$.  Hence, the time complexity of DIVISORS is
$O(N^{\epsilon}(\log N)^2)$.

Let $x \in G$ and $m=$ ord$(x)$. By  Lagrange's theorem \cite[page
35]{Ledermann}, $m$ divides $N$. Furthermore, by Lemma 1, if
 $m = p_1^{b_1}\cdots p_k^{b_k}$, then
for $i = 1,\ldots, k$ the element $x_i = (m/p_i^{b_i})x$ has order
$p_i^{b_i}$. The following algorithm computes correctly $m$ and
$m/p_i^{b_i}$ $(i=1,\ldots,k)$.

\ \\
{\bf ORDER }
\ \\
{  \sl Input:} $x\in G$ and $D(N)$.
\ \\
{ \sl Output:} $[m,m_1,\ldots,m_k]$, where ord$(x) = m =
p_1^{b_1}\cdots p_k^{b_k}$ and $m_i = m/p_i^{b_i}$.

\begin{enumerate}
\item For $l = 1,\ldots, \tau(N)$ \\
test wether or not  $d_lx  = 0$. If  $d_lx  = 0$, then   stop.\\
Put $m = d_l$.

\item For $i =1,\ldots, k$ compute  $m_i = m/p_i^{b_i}$.

\end{enumerate}

For the computation of $d_lx$, we need  $O({\log}N)$ operations in
$G$ \cite[page 69]{Gathen} and so,  the computation of all $d_l x$
needs $O(\tau(N) (\log N)^2)$. The computation of every $m_i$
needs $O((\log m) (\log p_{i}^{b_i}))$ bit operations and hence,
the computation  of all $m_i$ requires $O((\log N)^2)$ bit
operations. Therefore, the time complexity of the ORDER algorithm
is $O(N^{\epsilon}(\log N)^2)$.

 Suppose that $B= \{b_1,\ldots,b_n\}$ is a subset of $G$  such
that the group $H=<B>$ is the direct sum of the cyclic groups
$<b_i>$ $(i=1,\ldots,n)$. The {\em extending discrete logarithm
problem (EDLP)} is  the following problem: \\
Given a set $B \subseteq G$  as above and $w \in G$, determine the
smallest positive integer $z$ with $zw \in H$ and positive
integers $z_1, \ldots, z_n$ satisfying
$$zw = \sum_{i=1}^{n}z_i b_i.$$
Note that $z \leq  $ ord$(w)$. If $z = $ ord$(w)$, then $H\cap <w>
= \{0\}$. In \cite{Teske}, an algorithm is presented which solves
EDLP with running time
$$O( \max \{\lceil p^{1/2}\rceil^{n} e(p)\}),$$
where the maximum is taken over all prime divisors of $N$ and
$p^{e(p)}$ is the exponent of the $p$-component of $G$. It is
called SOLVE-EDLP.  Thus, we have SOLVE-EDLP$(w,B) =
(z,z_1,\ldots,z_n)$. On the other hand,  \cite{Shoup} implies that
in case where the order of $b_i$ is a power of a prime $p$, the
expression of a given element on a given basis of $H$ requires at
least $\Omega(p^{n/2})$ operations.

Let $p$ be a prime divisor of $N$ and $G(p)$ the $p$-component of
$G$. Suppose that $B = \{b_1,\ldots, b_n\}$ be a subset of $G(p)$
such that the group $H=<B>$ is the direct sum of the cyclic groups
$<b_i>$, $(i = 1,\ldots,n)$. If $x\in G(p)$, then we denote by
$H^{\star}$ the group generated by the set $B \cup \{x\}$. Suppose
that the orders of elements of $B$ are known and we have a
relation of the form
$$p^k x = \sum_{i=1}^{n} \delta_i b_i,$$
where $\delta_i \in  \Zset$ and $0 \leq \delta_i < $ ord$(b_i)$
$(i =1,\ldots,n)$. In \cite{Beynon}, an algorithm is given called
BASIS which computes a basis for $H^{\star}$ with running time
$O((\log |H^{\star}|)^2)$. If $B^{\star}$ is the basis of
$H^{\star}$ computed by BASIS, then we write
BASIS$(B,x,(p^k,\delta_1,\ldots,\delta_n))= B^{\star}$.

\section{Proof of Theorem 1}

In this section we develop an algorithm for finding a basis of a
finite abelian group $G$. If   $A$ and  $B$ are two subsets of
$G$, then we recall that $A\setminus B$ denotes the set of
elements of $x$ which do not belong to $B$. As we have noted in
the Introduction, for every prime $p$ dividing $|G|$, we have to
compute a basis for the $p$-primary component $G(p)$ of $G$.

\ \\
{\bf BASIS1}
\ \\
{ \sl Input:} An abelian group $(G,+)$ with  $|G| = N$ and
the prime factorization $N = p_1^{a_1}\cdots p_k^{a_k}$ of $N$.
\ \\
{\sl Output:} For $i=1,\ldots,k$, $(y_{1,i},n_{1,i}), \ldots ,
(y_{l(i),i},n_{l(i),i})$, where $y_{j,i}\in G$ with  ord$(y_{j,i})
= n_{j,i}$, such that  the $p_i$-primary component of $G$ is \\
$G(p_i) = <y_{1,i}> \oplus \cdots \oplus <y_{l(i),i}>$.

\begin{enumerate}
\item  Set $G_0(p_i) = \{0\}$, $B_{0,i} = \O$, $(i =
1,\ldots, k)$ and $G_0  = \{0\}$.

\item  Compute DIVISORS($N=p_{1}^{a_1} \cdots p_{k}^{a_k}$) =
$D(N)$.

\item For $j = 1,2,3 \ldots $
\begin{enumerate}

\item Choose  $x_j \in G  \setminus G_{j-1}$.

\item Compute ORDER$(x_j,D(N))= [m_j,m_{j,1},\ldots,m_{j,k}]$.
\item  For $i=1,\ldots,k$
\begin{enumerate}
\item Compute $x_{j,i} = m_{j,i} x_j$.

\item Compute
SOLVE-EDLP$(x_{j,i},B_{j-1,i}) =
(z_{j,i},z_{j,i,1},\ldots,z_{j,i,n})$.

\item Compute the bigger integer $k_{j,i}\geq 0$ such that
$p_{i}^{k_{j,i}}$ divides $z_{j,i}$, $s_{j,i} = z_{j,i}/p_{i}^{k_{j,i}}$
and $h_{j,i} = s_{j,i}x_{j,i}$.

\item Compute
 BASIS$(B_{j-1},h_{j,i},(p_{i}^{k_{j,i}},z_{j,i,1},\ldots,z_{j,i,n}))= B_{j,i}$.

\item Set $G_j(p_i) = <B_{j,i}>$ and  $G_j = G_j(p_1)+\cdots + G_j(p_k)$.
\end{enumerate}
\item Compute the elements of $G_{j} \setminus G_{j-1}$.

\item If $|G_j| = N$, then stop.
\end{enumerate}
\item  Output the couples  $(y_{1,i},n_{1,i}), \ldots ,
(y_{l(i),i},n_{l(i),i})$, where $y_{1,i},\ldots, y_{l(i),i}$ are
the elements of $B_{j,i}$ and ord$(y_{j,i}) = n_{j,i}$ $(i =
1,\ldots,k)$.

\end{enumerate}

{\bf Proof of correctness of BASIS1}

Since in every step $j$ we choose an element $x_j \in G\setminus
G_{j-1}$, at least one of the elements $x_{j,i}$ does not belongs
to $G_{j-1}$ and so $G_j$ is strictly bigger than $G_{j-1}$. For
 $j = 1,2,\ldots,$ BASIS1 constructs a basis for the group
$<\{x_{j,i}\}\cup B_{j-1,i}>$  and, after a finite number of
steps,  we obtain $j=r$ with $G_r(p_i) = G(p_i)$ $(i=1,\ldots,k)$.
Hence, the elements of the sets $B_{r,i}$ $(i=1,\ldots,k)$ form a
basis for $G(p_i)$. So, $G_r = G$ and  the elements of the sets
$B_{r,i}$ $(i=1,\ldots,k)$ form a basis for $G$.

{\bf Time Complexity of BASIS1}

Step 2 uses $O(N^{\epsilon}(\log N)^2)$ numerical operations.
Suppose that   $G_r = G$. A bound for  $r$  is given by $\tau(N)$,
the number of divisors of $Í$, and so $r = O(N^{\epsilon})$, where
$\epsilon$ is arbitrary small. In Step 3(b), we repeat $r$ times
the procedure ORDER and so, the time complexity of Step 3(b) is
$O(N^{\epsilon}(\log N)^2)$. The computation of each $x_{j,i}$
requires $O(\log  m_{j,i})$ group operations \cite[page
69]{Gathen}. Since $m_{j,i} \leq p_i^{a_i}$ the time complexity of
Step 3c(i) is
$$O(r \sum_{i=1}^{k} \log p_i^{a_i}) = O(N^{\epsilon} \log N)$$
group operations. Let $\mu(p_i)$ be the
 number of cyclic groups which are direct factors of $G(p_i)$.
The Step 3c(ii) uses the procedure SOLVE-EDLP and so, its time
complexity is
$$O( \sum_{i=1}^{r}\sum_{i=1}^{k}\lceil p_i^{1/2}\rceil^{|B_{j-1,i}|} e(p_i)) =
O(N^{\epsilon} \sum_{i=1}^{k}\lceil p_i^{1/2}\rceil^{\mu(p_i)}
e(p_i)) = O(N).$$ The Step 3c(iii) has time complexity
$$O(r\sum_{i=1}^{k} (\log s_{j,i}+a_i)) = O(N^{\epsilon} \log N).$$
The Step 3c(iv) uses the procedure BASIS and so, its time
complexity is
$$O(\sum_{j=1}^{r} \sum_{i=1}^{k}(\log |G_j(p_i)|)^2)=O(N^{\epsilon}(\log N)^2).$$
Since $|G_{j-1}| < |G_j|$ and $|G_{j-1}|$ divides $|G_j|$, we have
$\sum_{j=1}^r|G_j| \leq 2N$, and so, the time complexity of
Step3(d) is $O(N)$. Therefore, the time complexity of BASIS1 is
$O(N)$.

\section{Proof of Theorem 2}

In this section we develop our algorithm for finding a basis of a
finite abelian group $G$ in case where a generating system of $G$
is known.  We denote by $\mu(p)$ the number of cyclic subgroups
which are direct factors of $G(p)$ and by $p^{e(p)}$ the exponent
of $G(p)$.

\ \\
{\bf BASIS2}
\ \\
{\sl Input:} An abelian group $(G,+)$ with  $|G| = N$, a
generating system $\{g_1,\ldots,g_M\}$ for $G$ and the prime
factorization $N = p_1^{a_1}\cdots p_k^{a_k}$ of $N$. \\
{\sl Output:} For $i=1,\ldots,k$, $(y_{1,i},n_{1,i}), \ldots ,
(y_{l(i),i},n_{l(i),i})$, where $y_{j,i}\in G$ with  ord$(y_{j,i})
= n_{j,i}$, such that  the $p_i$-primary component of $G$ is \\
$G(p_i) = <y_{1,i}>\oplus \cdots \oplus <y_{l(i),i}>$.

\begin{enumerate}

\item  Compute DIVISORS($N=p_{1}^{a_1} \cdots p_{k}^{a_k}$) =
$D(N)$.

\item For $j = 1,\ldots, M$, compute ORDER$(g_j, D(N))= [m_j,m_{j,1},\ldots,m_{j,k}]$.

\item For  $i = 1,\ldots, k$ and $j = 1,\ldots, M$, compute
$g_{j,i} = m_{j,i} g_j$.

\item For $i = 1,\dots, k$,\\
Set $B_{1,i} = \{g_{1,i}\}$. \\
For  $j = 2,\ldots, M$, \\
If $|<B_{j,i}>| \neq p_i^{a_i}$,

\begin{enumerate}
\item Compute SOLVE-EDLP$(g_{j,i},B_{j-1,i}) =
(z_{j,i},z_{j,i,1},\ldots,z_{j,i,n})$.
\item  Compute the bigger integer $k_{j,i}\geq 0$ such that
$p_i^{k_{j,i}}$ divides $z_{j,i}$, $s_{j,i} =
z_{j,i}/p_i^{k_{j,i}}$ and $h_{j,i} = s_{j,i}g_{j,i}$.
\item Compute BASIS$(B_{j-1},h_{j,i},(p_i^{k_{j,i}},z_{j,i,1},\ldots,z_{j,i,n}))= B_{j,i}$.
\end{enumerate}
\item Output the couples  $(y_{1,i},n_{1,i}), \ldots ,
(y_{l(i),i},n_{l(i),i})$, where $y_{1,i},\ldots, y_{l(i),i}$ are
the elements of $B_{M,i}$ and ord$(y_{j,i}) = n_{j,i}$ $(i =
1,\ldots,k)$.
\end{enumerate}

{\bf Proof of correctness of BASIS2}

For $j = 1,\ldots M$ the algorithm constructs a basis of the group
$<g_{j,i},B_{j-1,i}>$ until a basis of $G(p_i)$ is obtained.

{\bf Time Complexity of BASIS2}

Step 1 requires  $O(N^{\epsilon}(\log N)^2)$ bit operations. The
complexity of Step 2 is $O(MN^{\epsilon}(\log N)^2)$, where
$\epsilon$ is arbitrary small. For the computation of every
$g_{j,i} = m_{j,i} g_j$ we need $O({\log}m_{j,i})$ group
operations. Since $m_{j,i} \leq p_{i}^{a_i}$, Step 3 requires $O(M
\log N)$ group operations. For $i = 1,\dots, k$,
  the use of SOLVE-EDLP, in Step 4(a), requires
$O(e(p_i) M\lceil p^{1/2}\rceil^{\mu(p_i)})$ operations, Step
4(b) $O(M \log p_i^{a_i})$ operations and the use of BASIS, in
Step 4(c), $O(M (\log p_i^{a_i})^2)$ operations. Hence, the time
complexity of  BASIS2 is
$$O(M\sum_{i=1}^{k} e(p_i)\lceil p^{1/2}\rceil^{\mu(p_i)}).$$
Moreover, using the lower bound for the time complexity of
SOLVE-EDLP, we have
$$\Omega(\sum_{i=1}^{k} p_i^{(\mu(p_i)-1)/2}).$$

\section{Proof of Theorem 3}

In this section, we suppose that  the abelian group $G$ is cyclic. We propose a simple algorithm
in case where a generating system of $G$ is known.

\ \\
{\bf BASIS3}
\ \\
{\sl Input:} A cyclic group $(G,+)$ with  $|G| = N$, a
generating system $\{g_1,\ldots,g_M\}$ for $G$ and the prime
factorization $N = p_1^{a_1}\cdots p_k^{a_k}$ of $N$. \\
{\sl Output:} $(y_1, \ldots , y_k)$,
where $y_{i} \in G$ with  ord$(y_{i}) = p_{i}^{a_i}$, and
so,  the $p_i$-primary component of $G$ is
$G(p_i) = <y_{i}>$.

\begin{enumerate}

\item  Compute DIVISORS($N=p_{1}^{a_1} \cdots p_{k}^{a_k}$) =
$D(N)$.

\item For $j = 1,\ldots, M$, compute ORDER$(g_j, D(N))= [m_j,m_{j,1},\ldots,m_{j,k}]$.

\item For  $i = 1,\ldots, k$ and $j = 1,\ldots, M$, compute
$g_{j,i} = m_{j,i} g_j$.

\item For $i = 1,\dots, k$, find $s \in \{1, \ldots, M \}$, with ord($g_{s,i}$) = $\max_{1\leq t \leq M }$ \{ord($g_{t, i}$)\},
and set $y_i = g_{s,i}$.
\item Output $(y_{1}, \ldots, y_{k})$.
\end{enumerate}

{\bf Proof of correctness of BASIS3}

We remark that $G(p_i)=<g_{1,i}, \ldots, g_{M,i}>$. On the other
hand,  since $G$ is a cyclic group, we have  $G(p_i) \cong
\Zset_{p_{i}^{a_i}}$. It follows that, the element having the
maximum order among $g_{1,i}, \ldots, g_{M,i}$, is the generator
of $G(p_i)$.

{\bf Time Complexity of BASIS3}

The Step 1 and 2, require $O(N^\epsilon (\log N)^2)$ and $O(M
N^\epsilon (\log N)^2)$ group operations, respectively. The Step 3
needs $O(M \log N)$ group operations. The Step 4 needs $O(kM)$
operations, and since  $k =O({\log \log}N)$ \cite[ page
359]{Hardy}, it follows that the time complexity of this step is
$O(M \log \log N)$ operations. Therefore, the time complexity of
BASIS3, is $O(M N^\epsilon)$.

{\bf Acknowledgments} \\
The first author gratefully acknowledges support of the project
Autonomic Network Arhitecture (ANA), under contract number
IST-27489, which is funded by the IST FET Program of the European
Commision.

\end{document}